\documentclass[pra,superscriptaddress,reprint]{revtex4-1}
\usepackage{graphicx}
\usepackage{enumitem}
\usepackage{amsmath}
\newcommand{\ket}[1]{|{#1}\rangle}
\newcommand{\bra}[1]{\langle{#1}|}

\usepackage{amsfonts}
\usepackage{bbold}

\usepackage[usenames,dvipsnames]{xcolor}
\usepackage[colorlinks=true,citecolor=Cerulean,linkcolor=RubineRed,urlcolor=Cerulean]{hyperref}

\def\PKS{Max-Planck-Institut f\"{u}r Physik komplexer Systeme, D-01187 Dresden, Germany}
\def\SUPA{SUPA, Institute of Photonics and Quantum Sciences, Heriot-Watt University, Edinburgh, EH14 4AS, UK}
\def\AU{Department of Physics and Astronomy, Aarhus University, DK-8000 Aarhus C, Denmark}

\begin{document}

\title{Topological models in rotationally symmetric quasicrystals}

\author{Callum W. Duncan}
\email{duncan@pks.mpg.de}
\affiliation{\PKS}
\affiliation{\SUPA}

\author{Sourav Manna}
\affiliation{\PKS}

\author{Anne E. B. Nielsen}
\altaffiliation{On leave from \AU}
\affiliation{\PKS}

\begin{abstract}
We investigate the physics of quasicrystalline models in the presence of a uniform magnetic field, focusing on the presence and construction of topological states. This is done by using the Hofstadter model but with the sites and couplings denoted by the vertex model of the quasicrystal, giving the Hofstadter vertex model. We specifically consider two-dimensional quasicrystals made from tilings of two tiles with incommensurate areas, focusing on the five-fold Penrose and the eight-fold Ammann-Beenker tilings. This introduces two competing scales; the uniform magnetic field and the incommensurate scale of the cells of the tiling. Due to these competing scales the periodicity of the Hofstadter butterfly is destroyed. We observe the presence of topological edge states on the boundary of the system via the Bott index that exhibit two way transport along the edge. For the eight-fold tiling we also observe internal edge-like states with non-zero Bott index, which exhibit two way transport along this internal edge. The presence of these internal edge states is a new characteristic of quasicrystalline models in magnetic fields. We then move on to considering interacting systems. This is challenging, in part because exact diagonalization on a few tens of sites is not expected to be enough to accurately capture the physics of the quasicrystalline system, and in part because it is not clear how to construct topological flatbands having a large number of states. We show that these problems can be circumvented by building the models analytically, and in this way we construct models with Laughlin type fractional quantum Hall ground states.
\end{abstract}

\maketitle

\section{Introduction}

The emergence of the integer \cite{Klitzing1980} and fractional \cite{Laughlin1981,Tsui1982} quantum Hall effects led to the discovery of a new set of phases that are beyond Landau's theory of symmetry breaking \cite{Landau1958}. This resulted in the emergence of a new theme in condensed matter physics associated to new states of matter due to topological order \cite{Hasan2010,Wen2017}. The need to go beyond Landau's theory of symmetry breaking is shown by the fact that the different phases arising have the same symmetry \cite{Wen1990}. Therefore, no local order parameter can distinguish them. The quantity that distinguishes them is topological and is characterised by a global order parameter \cite{Wen2017,Hamma2008}. A consequence of topological order being present in a system is the appearance of quasiparticle excitations in two dimensions with fractional statistics called anyons \cite{lerda2008}.

There is currently also a large amount of interest in topological insulators and Chern insulators, which have gapless conducting states at boundaries where the topological invariant changes \cite{Hasan2010}. These states are commonly studied by either the calculation of topological invariants \cite{Thouless1982,Kitaev2009} or by the direct calculation of the edge states \cite{Duncan2018,Pantale2017,Kunst2017}. The generality of topological systems is shown by the variety of experimental systems where they have been realised, including condensed matter \cite{hsieh2008,xia2009}, ultracold atoms \cite{atala2013,jotzu2014,meier2016}, photonic lattices \cite{mukherjee2017,maczewsky2017}, acoustics \cite{he2016} and mechanical systems \cite{Susstrunk2015}.

The idea that ground states of matter are always periodic crystals began to weaken in the 1960s \cite{janssen2018}, first with the observation of modulated phases \cite{Brouns1964,Wolff1974}, then with the discovery of quasicrystals by Schechtman \cite{Shechtman1984}. One method to model quasicrystals is to consider atoms located at the vertices of aperiodic tilings, these are called vertex models. There has been a variety of works on modelling quasicrystals in such a fashion \cite{Tsunetsugu1991,Rieth1995,Repetowicz1998,Prunel2002}. Note, that a lattice is defined as being periodic. We will therefore refer to the collection of sites of the quasicrystalline models as a quasilattice throughout this work, as is done in Ref.~\cite{loring2019bulk}.

Recently, there has been some interest in the prescence of topological states in quasicrystals, including for fermions with spin \cite{chen2019,varjas2019}, photonic lattices \cite{Bandres2016}, maxwell lattices \cite{Zhou2019}, vortex gap solitons \cite{Sakaguchi2006}, superconductors \cite{Fulga2016} and Hofstadter models \cite{Tran2015,Fuchs2016}.  Note, the concept of bands and gaps in the spectrum are not well-defined in quasicrystals and is currently an open problem \cite{Tsunetsugu1991,Naumis1994}. In this work, we will consider the states directly, observing the properties expected of topological states in regular lattices. The previous works on the vertex model for a quasicrystal with a magnetic field consider the Rauzy tiling. The isomorphism of the Rauzy tiling considered in these works is specifically constructed to have the two-dimensional tiling constructed of a single tile \cite{Vidal2001,Fuchs2016,Tran2015}. However, in other quasicrystalline systems the incommensurate nature of the multiple tiles is a common property with physical consequences. The Rauzy tiling, due to its single tile, shows similar physics to the periodic two-dimensional lattice in the presence of a magnetic field, e.g.\ a periodic Hofstadter butterfly and edge states.

In this work, we consider both five-fold and eight-fold quasicrystalline models with a perpendicular magnetic field applied to them. Here we are specifically interested in retaining the quasiperiodic nature of the different tiles inherent to most quasicrystals and exploring its consequence for the physics of the system. The importance of the incommensurate nature of the tiling will be shown by the eigenvalues and eigenstates of the system as studied in Sec.~\ref{sec:NonInter}. In order to show the importance of the incommensurate fluxes in each cell of the tiling, we will also compare the quasicrystal results to that of a periodic square lattice with two alternating incommensurate fluxes. We then study the topological edge states in detail and characterise them by using the Bott index. We observe the surprising feature that there are also internal edge states in the eight-fold Ammann-Beenker tiling and corresponding transport along an internal edge. Investigating the spectrum, we find topologically non-trivial nearly flatbands, but only with a relatively small number of states.

If bands with a non-zero Chern number are made flat, they are similar to Landau levels. It has been demonstrated for several models on regular two-dimensional lattices that one can produce topologically non-trivial flatbands and obtain fractional quantum Hall type physics by adding interactions, see e.g.\ \cite{Neupert2011,regnault2011}. This approach seems difficult to use in quasicrystalline quasilattices. Due to the irregularity of the tiling and lack of periodicity, it is not clear how to obtain a topologically non-trivial flatband with a large number of states, and even if one found such a band, it would not be easy to check the properties after adding interactions, since quite large systems are needed to appropriately capture the physics of the quasilattices \cite{loring2019bulk}. Another approach to obtain fractional quantum Hall physics on lattices is to start from a trial fractional quantum Hall state, modify it to be defined on the lattice under consideration, and then analytically derive a parent Hamiltonian for the state \cite{nielsen2012,tu2014}. This approach is quite insensitive to the details of the lattice structure \cite{glasser2016}, and in Sec.\ \ref{sec:Inter} we use it to construct Laughlin type fractional quantum Hall models on quasicrystalline models with five-fold Penrose and eight-fold Ammann-Beenker tilings.

\section{Quasicrystalline quasilattices from tiling}

Quasicrystalline structures can be constructed in a variety of ways, including from non-periodic tilings and from the interference of periodic lattice structures. In this work we will consider five- and eight-fold symmetric quasilattices. We will characterise the form of the quasilattices from non-periodic tilings and utilise the triangle decomposition method for the five-fold tiling and the cut-and-project method for the eight-fold tiling. Throughout this work we will consider quasilattices with open boundary conditions.

\subsection{The five-fold quasilattice}

The Penrose tiling is an example of an aperiodic set of prototiles with five-fold rotational symmetry. As the Penrose tiling is aperiodic, it lacks translational symmetry but it is self-similar at large scales. There are various methods to generate Penrose tilings of different forms, including via two-dimensional projections of five-dimensional cubic structures \cite{DEBRUIJN1981}. We will form Penrose tilings via the Robinson triangle decomposition method, which gives the Penrose tiling with rhombuses \cite{walter2009}. This starts with two triangles of different sizes (an A-type and B-type) which when fused together generate the two rhombuses of the Penrose tiling. The triangles have defined apex angles of $\theta_A = 36^\circ$ and $\theta_B=108^\circ$. To generate the tiling pattern the following steps are utilised: (1) Start with a single one of the Robinson triangles (A or B). (2) Decompose this triangle into two Robinson triangles. (3) Repeat this decomposition for all triangles a number of times. After this the Penrose tiling of rhombuses is obtained by fusing triangles of the same type and touching bases. The size of each tile is dependent on the starting size of the triangle and the number of decompositions. However, all tiles will have sides of length $l$.

\begin{figure}[t]
	\centering
	\includegraphics[width=0.98\linewidth]{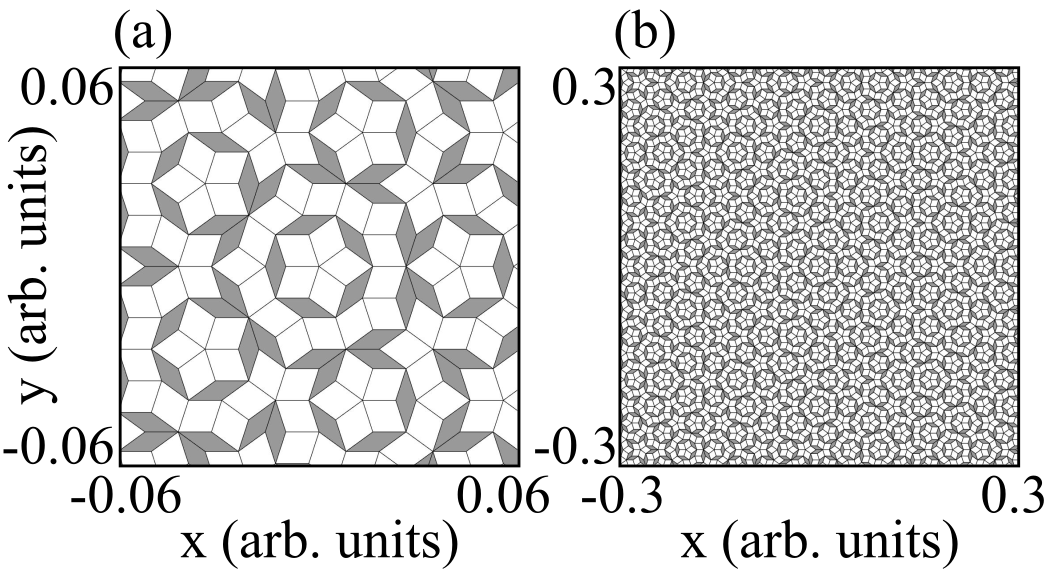}
	\caption{Examples of the rhombus Penrose tiling obtained from the Robinson triangle decomposition method. (a) A zoomed in portion of the Penrose tiling showing the two types of rhombuses by shading. (b) A larger portion of the tiling showing the self-similar structure.}
	\label{fig:PenroseTile}
\end{figure}

\begin{figure}[t]
	\centering
	\includegraphics[width=0.98\linewidth]{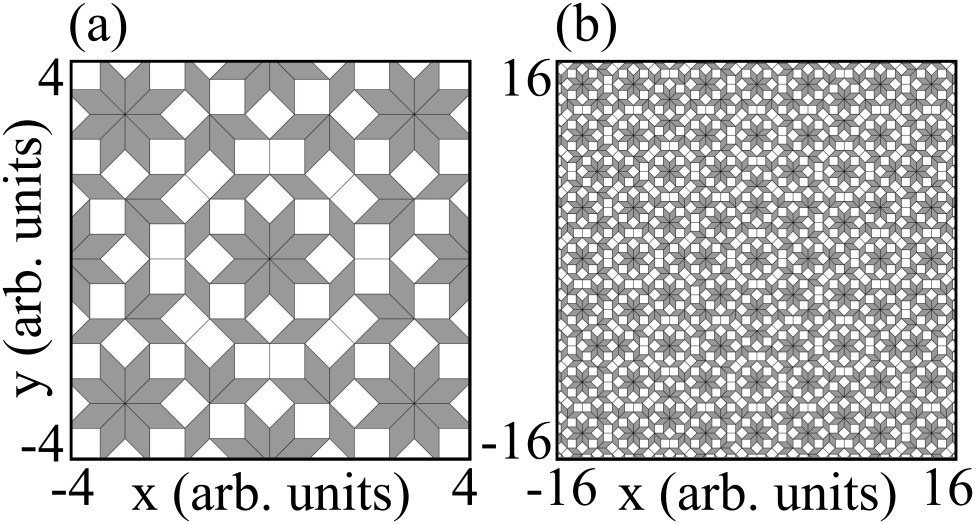}
	\caption{Examples of the rhombus Ammann-Beenker tiling obtained from the cut-and-project method. (a) A zoomed in portion of the Ammann-Beenker tiling showing the two types of rhombuses by shading. (b) A larger portion of the tiling showing the self-similar structure.}
	\label{fig:AmmBeenLatt}
\end{figure}

In Fig.~\ref{fig:PenroseTile}, we show the Penrose tiling obtained from the Robinson triangle decomposition method at a small and large scale. The tiling shown is obtained for $10$ decompositions of the $A$ triangle and will be used throughout this work. The five-fold symmetry in the quasilattice is clear in both the small and large scale figures.

\subsection{The eight-fold quasilattice}

The Ammann-Beenker tiling is an example of an aperiodic set of prototiles with eight-fold rotational symmetry. Again, this tiling lacks translational symmetry but does have a self-similar structure at large scales. There are also various methods to generate the eight-fold tiling but we will utilise the cut-and-project method \cite{moody2000,baake2002,baake2013}. The cut-and-project method is simplest to explain in the context of obtaining a one-dimensional non-periodic tight-binding model. In that scenario, the non-periodic one-dimensional model is obtained from a periodic two-dimensional structure. First, a line with irrational slope is drawn through the periodic two-dimensional lattice and then the two-dimensional lattice points are projected onto this line to obtain a one-dimensional non-periodic model. In order to limit the number of sites for the one-dimensional model it is possible to project only points within a certain range of the line. To extend this to form two-dimensional non-periodic quasilattices is straight-forward with the consideration of an irrational plane in a higher dimensional space. Note, this higher dimensional space is a superspace and it was recently utilised to develop a Hamiltonian formalism in superspace to obtain the eigenstates of a quaisperiodic model \cite{valiente2019}.

For the specific case of the eight-fold quasilattice and the Ammann-Beenker tiling the higher dimensional space is that of a four-dimensional cubic lattice that has an eight-fold symmetry. A two-dimensional irrational plane is selected and the lattice sites that are within an octagon, of a chosen size, surrounding this plane are projected onto the plane \cite{varjas2019}. Note, the choice of the shape around the plane within which the four-dimensional lattice sites are projected is solely responsible for the number of sites obtained in the two-dimensional model and the form of the boundary. We will consider in this work boundaries that preserve the special symmetries of the quasilattice. The tiles of the Ammann-Beenker tiling consist of a square and a rhombus defined with angles $135^\circ$ and $45^\circ$. Again, all tiles have sides with a length of $l$.

In Fig.~\ref{fig:AmmBeenLatt}, we show the Ammann-Beenker tiling obtained from the cut-and-project method at a small and large scale. Again, the eight-fold symmetry in the quasilattice is clear in both the small and large scale figures.

\section{The Hofstadter vertex model}
\label{sec:NonInter}

We will in this section consider the vertex model, which assumes atoms sitting at the vertices of the tiling and the bonds are given by the connections along the rhombuses. The standard vertex model has a Hamiltonian of the form
\begin{equation}
H = \sum_{\langle i,j \rangle} t_{ij} \ket{i} \bra{j},
\label{eq:VertexHam}
\end{equation}
with $\ket{i}$ the wave function at site $i$ and $t_{ij}$ is the tunnelling strength between sites. The summation is over all sites that are connected along the side of one of the rhombuses. The physics of the Hamiltonian \eqref{eq:VertexHam} with constant and real $t_{ij}$ is well known. Results for the five-fold case show that the spectrum has many gaps below and above zero energy due to the non-periodic nature of the quasicrystal \cite{Kohmoto1986}. There is also a number of states in a flat band at zero energy with localised states, which is a well-known characteristic property of the vertex models, including in a magnetic field \cite{Kohmoto1986b,Rieth1995,Arai1988}.

\begin{figure}[t]
	\centering
	\includegraphics[width=0.98\linewidth]{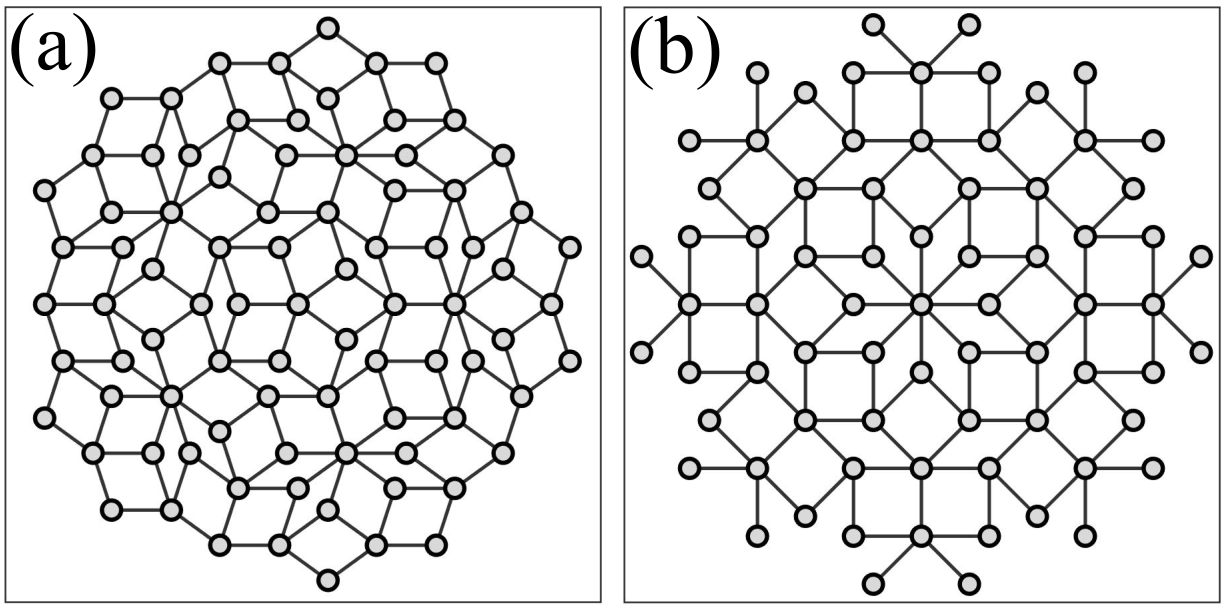}
	\caption{Examples of the quasilattice structures considered in this work. Solid lines show the couplings of the vertex model. (a) The five-fold symmetric quasilattice of the vertex model on the Penrose tiling. Shown is an example with $86$ sites. (b) The eight-fold symmetric quasilattice of the vertex model on the Ammann-Beenker tiling. Shown is an example with $73$ sites.}
	\label{fig:Lattices}
\end{figure}

As already stated, we will consider quasicrystal quasilattices with open boundary conditions which have five- and eight-fold rotational symmetries. In Fig.~\ref{fig:Lattices}, we show two examples of the quasilattices considered throughout this work. We show small quasilattices to allow for the simple visualisation of the couplings and symmetries.

We will consider the effect of a magnetic field being applied perpendicularly to the quasicrystal. It is well-known that this introduces a Peierls phase factor \cite{Hofstadter1976} when tunnelling between sites. Therefore, the tight-binding Hamiltonian takes the form
\begin{equation}
H = -J \sum_{\langle i,j \rangle} e^{i\theta_{ij}} \hat{b}_j^\dagger \hat{b}_i,
\label{eq:Hofs}
\end{equation}
with $\hat{b}_i^\dagger$ ($ \hat{b}_i$) the creation (annihilation) operator at the $i$th site. The summation is the same as in the standard vertex model. The phase is given by the line integral of the vector potential
\begin{equation}
\theta_{ij} = \int_{C_{ij}} \mathbf{A}(\mathbf{r}) \cdot d\mathbf{r},
\end{equation}
with $\mathbf{r}$ the space coordinate, $\mathbf{A}(\mathbf{r})$ the vector potential and $C_{ij}$ the path from site $i$ to site $j$. Throughout this work we will consider units of $\hbar = e = 1$, with the flux quantum $\phi_0 = 2 \pi$. We will take the Landau gauge, which has a vector potential of
\begin{equation}
\mathbf{A}(\mathbf{r}) = B \left(0,x,0 \right),
\end{equation}
with $B$ the magnetic field strength.

A major difference between a regular periodic lattice and a quasicrystal is the ability to define a repeated unit cell. For the quasicrystal such a unit cell cannot be defined. When there are multiple incommensurate area tiles this means the flux is not commensurate through each.

The phase introduced in the Landau gauge between two generic sites $(x_i,y_i)$ and $(x_j,y_j)$ is given by
\begin{equation}
\theta_{ij} = \frac{B}{2}(y_j - y_i) \left(x_i + x_j\right),
\end{equation}
for a detailed calculation see Appendix~\ref{app:GeneralPhase}. For the case of the sites existing on a Penrose tiling the phase factor of the tunnelling in the Landau gauge can be written as
\begin{equation}
\theta_{ij}^{5} = \phi \frac{\left(y_j - y_i\right) \left(x_i + x_j\right)}{2 l^2 \sin\theta_B} ,
\end{equation}
where $\phi$ is the flux penetrating the fat rhombus, i.e.\ $\phi = B l^2 \sin \theta_B$. We can similarly define this for the Ammann-Beenker tiling to have a phase factor of
\begin{equation}
\theta_{ij}^8 = \phi \frac{\left(y_j - y_i\right) \left(x_i + x_j\right)}{2 l^2} ,
\end{equation}
here $\phi$ is the flux penetrating the square, i.e.\ $\phi = B l^2$. Throughout this manuscript we will work in units of length $l$, that is the bond length between vertex sites.

\begin{figure}[t]
	\centering
	\includegraphics[width=0.95\linewidth]{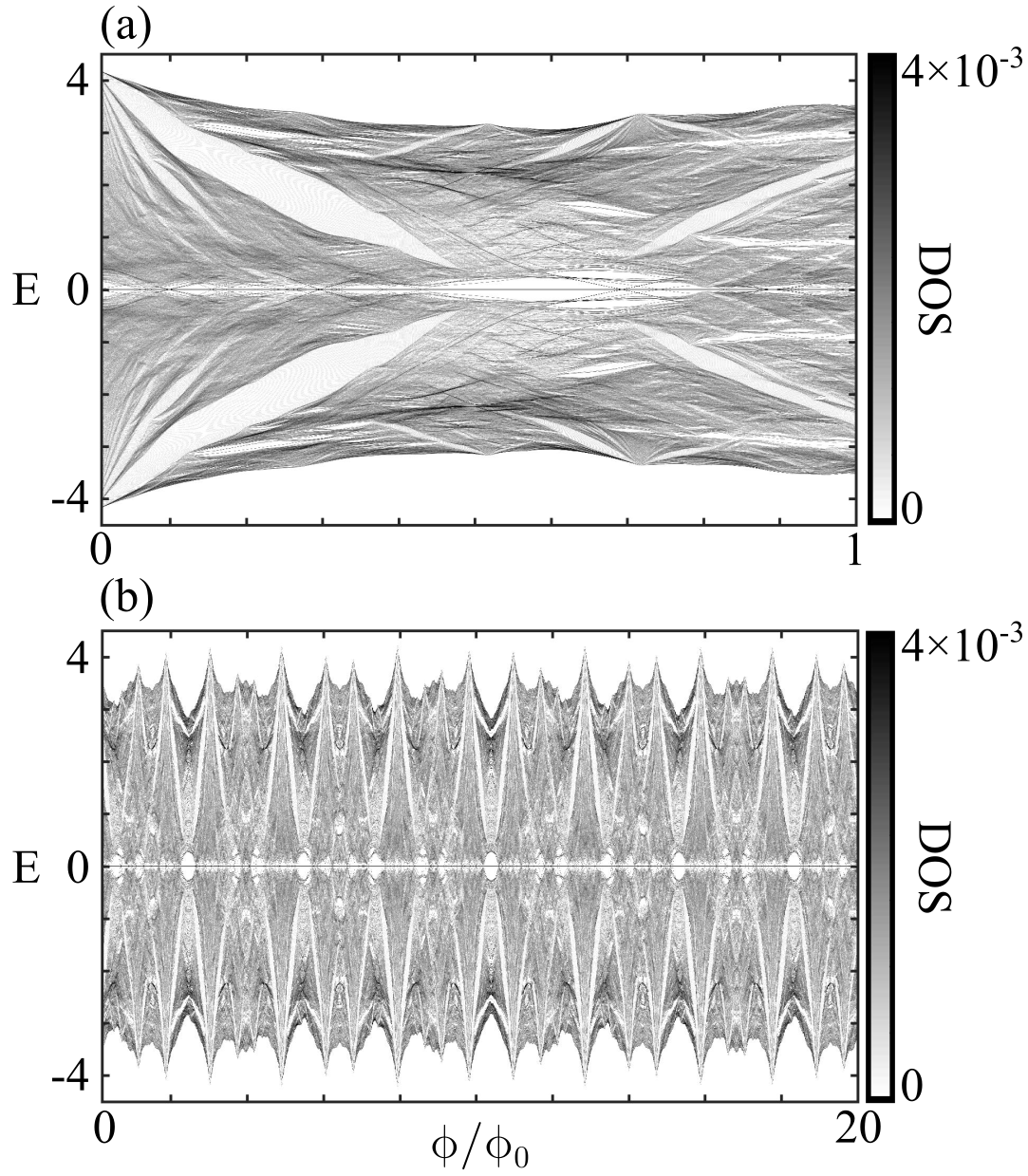}
	\caption{Energy-flux plane for the Hofstadter vertex model on the five-fold Penrose tiling with $3448$ sites. (a) The standard flux 0 to 1 plot. (b) A larger region of flux showing the incommensurate nature in the energy-flux plane.}
	\label{fig:HofsButt}
\end{figure}

\begin{figure}[t]
	\centering
	\includegraphics[width=0.95\linewidth]{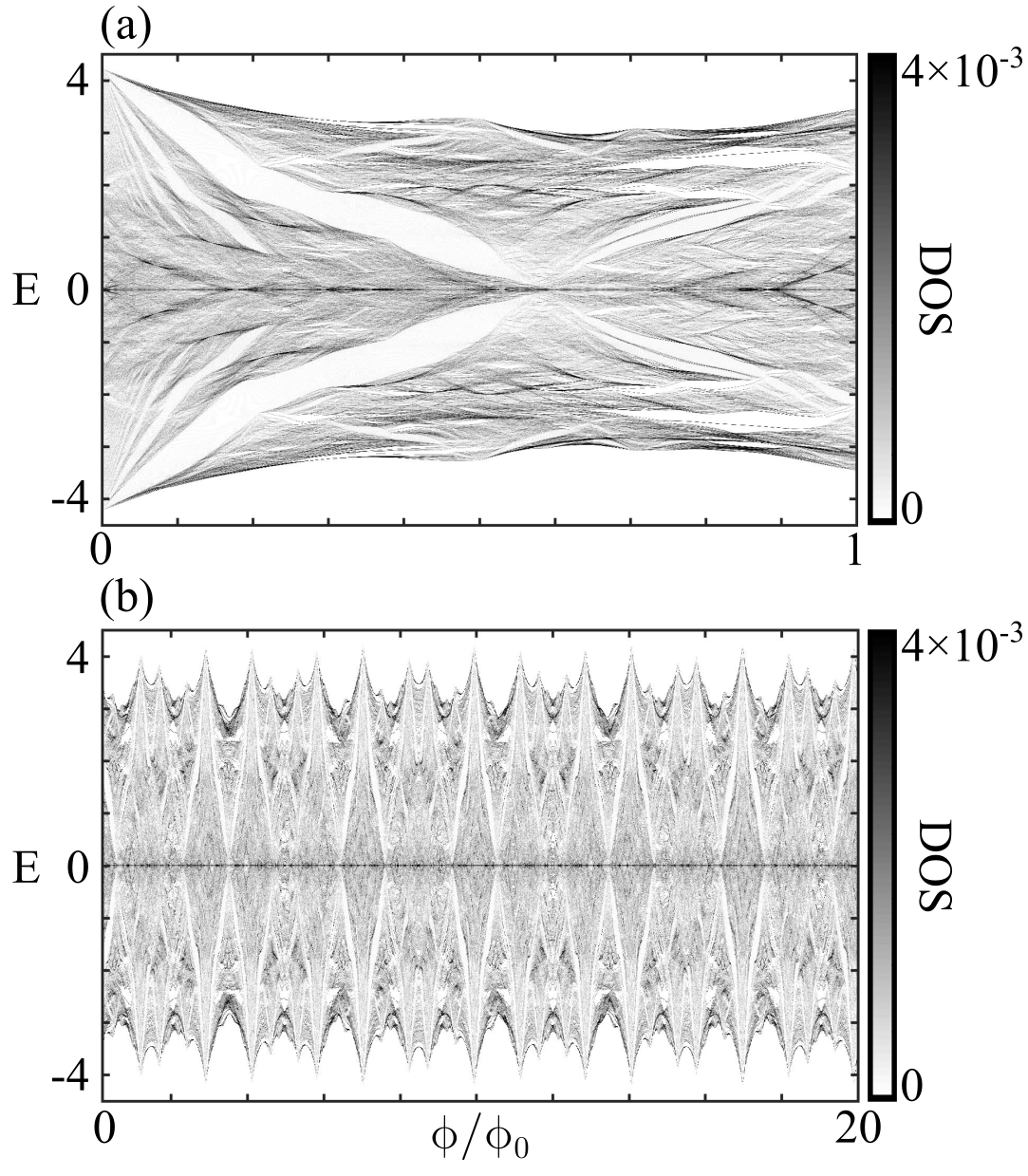}
	\caption{Energy-flux plane for the Hofstadter vertex model on the eight-fold Ammann-Beenker tiling with $3345$ sites. (a) The standard flux 0 to 1 plot. (b) A larger region of flux showing the incommensurate nature in the energy-flux plane.}
	\label{fig:HofsButt8Fold}
\end{figure}

\subsection{Hofstadter butterfly}
\label{sec:HofsButt}

For the standard two-dimensional square lattice, the presence of a magnetic flux opens up gaps in the spectrum. This leads to a fractal structure in the energy-flux plane called the Hofstadter butterfly \cite{Hofstadter1976}. The gaps that are opened usually lead to edge states with associated non-zero Chern numbers \cite{Hatsugai1993}. The structure of the Hofstadter butterfly in vertex models of quasicrystals with magnetic fields has been previously studied \cite{Hatakeyama1989,grimm1998,Fuchs2016,Naumis2019,grimm2011}. We will consider the Hofstadter butterflies briefly in this section and compare them to a regular periodic lattice with incommensurate staggered fluxes, which is a new approach.

The quasicrystal structure considered in this work is not constructed from a single cell, and this will have consequences for the structure in the energy-flux plane. One immediate consequence is that the energy-flux plane will no longer be periodic in regions of $0 < \phi/\phi_0 \leq 1$.

We consider the energy-flux plane of the quasicrystal in Fig.~\ref{fig:HofsButt} for $3448$ sites on the Penrose tiling. We are working with a large number of eigenstates, therefore the opening of gaps is easier to observe by using the density of states (DOS). As expected, we observe that the structure is no longer periodic. Instead we observe a more complex structure but still with the characteristic low density regions of the Hofstadter model. If we consider the standard $0 < \phi/\phi_0 \leq 1$ for the flux, as shown in Fig.~\ref{fig:HofsButt}(a), then we see what looks like the main gaps of the Hofstadter butterfly for $\phi/\phi_0 < 0.5$ and the usual smaller gaps at smaller $\phi$. However, we do not observe the DOS to be symmetric around $\phi/\phi_0=0.5$. Instead, there are smaller gapped regions and intricate structures shown. We also consider the energy-flux plane for an eight-fold quasilattice on the Ammann-Beenker tiling with $3345$ sites in Fig.~\ref{fig:HofsButt8Fold}. We observe a similar structure to the energy-flux plane as to the five-fold case, with the main gaps preserved for $\phi/\phi_0<0.5$ in Fig.~\ref{fig:HofsButt8Fold}(a).

When we look at a larger flux region, see Fig.~\ref{fig:HofsButt}(b) and Fig.~\ref{fig:HofsButt8Fold}(b), we observe an aperiodic nature of the energy-flux plane as was found in early works on magnetic fields in Penrose tilings \cite{Hatakeyama1989} and for the Hall voltage in the Ammann-Beenker tiling \cite{grimm1998}. The complex structure of the energy-flux plane is due to the two different area rhombuses for both symmetries and the natural non-periodic nature of a quasicrystal. The two different cells of the tiling introduce two length scales to the particle's phase under the influence of the magnetic field. We can investigate this statement by considering a regular square lattice geometry of the standard Hofstadter model.

\begin{figure}[t]
	\centering
	\includegraphics[width=0.65\linewidth]{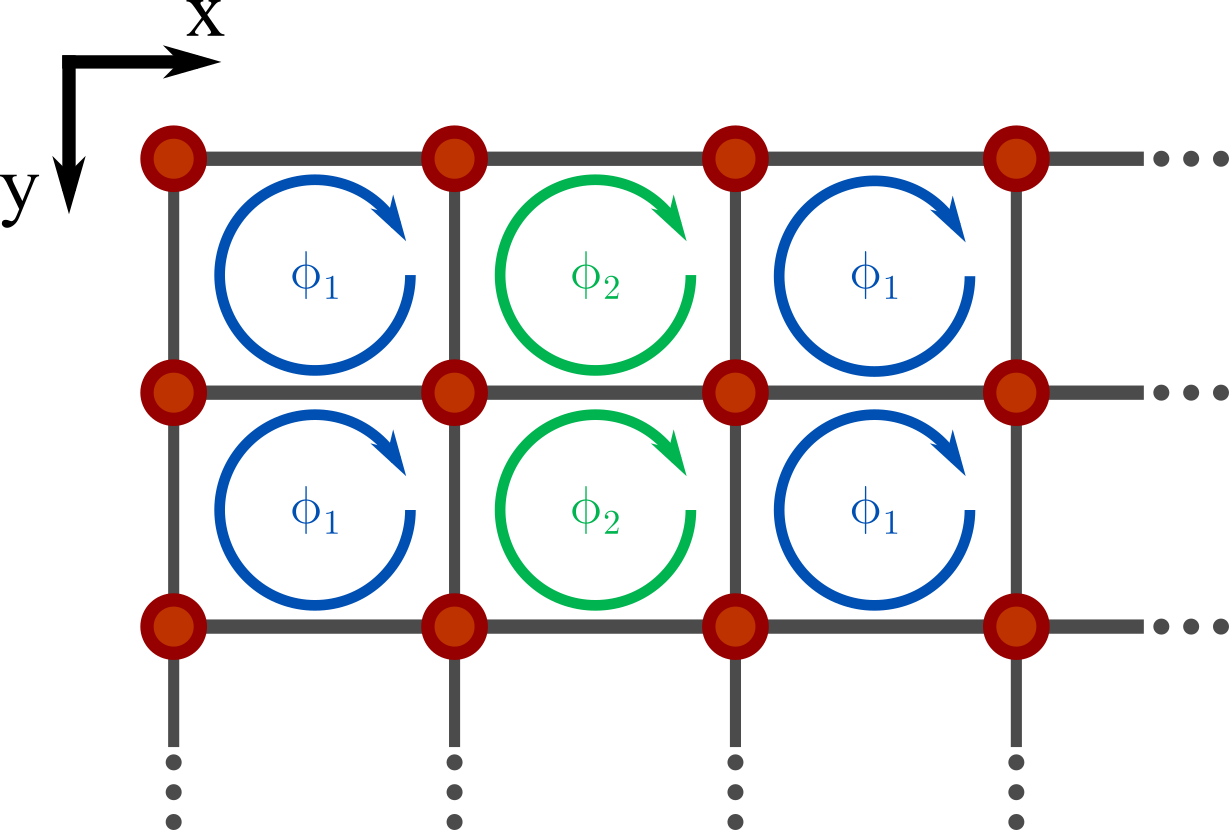}
	\caption{Illustration of the staggered flux square lattice Hofstadter model considered, with $\phi_1$ and $\phi_2$ two different fluxes that are not necessarily commensurate with each other.}
	\label{fig:HofsSquare}
\end{figure}

\begin{figure}[t]
	\centering
	\includegraphics[width=0.98\linewidth]{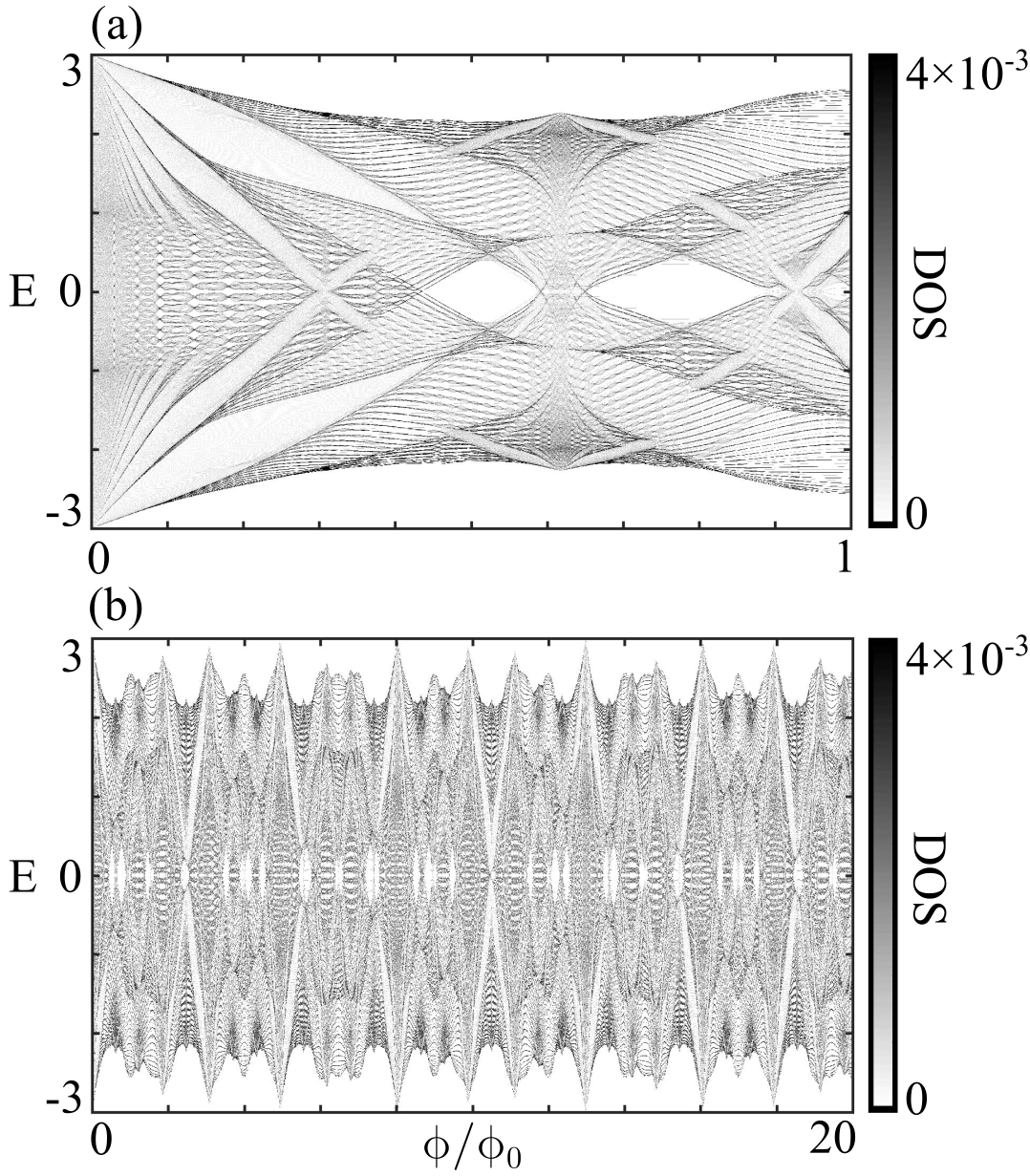}
	\caption{Energy-flux plane for the Hofstadter vertex model on the square lattice with incommensurate cells  and $1600$ sites with $\phi_1 = \tau$ and $\phi_2 = 1$. (a) The standard flux 0 to 1 plot. (b) A larger region of flux showing the incommensurate nature in the energy-flux plane.}
	\label{fig:HofsButtSquare}
\end{figure}

In the standard Hofstadter model the Hamiltonian of the system can still be described by Eq.~\eqref{eq:Hofs}, but now with the vertices being on a regular periodic square tiling. We can consider the influence of different unit cells for the lattice and magnetic field rather simply in the square lattice. We can take the flux in each unit cell of the lattice to be constant for translation in $x$ but staggered in $y$, see Fig.~\ref{fig:HofsSquare} for an illustration of this lattice. This new toy model allows the importance of different size cells to be probed. By setting the area of each cell the flux gained upon a circulation of that cell will change. If we set the two staggering fluxes to commensurate quantities, e.g.\ $\phi_1 = \phi$ and $\phi_2 = 2\phi$, then we get interesting structures for small flux regions that repeat themselves over a longer period than the standard Hofstadter butterfly. However, if we consider the two fluxes to be incommensurate then we would expect the periodic nature of the Hofstadter butterfly to be destroyed.

For the two incommensurate area cells of the square lattice we will consider the area of each to be $\phi_1=\tau$ and $\phi_2=1$, with $\tau$ the golden ratio. The ratio of the areas of the two cells is therefore the same as the case of the five-fold Penrose quasilattice considered, with the eight-fold quasilattice having a ratio of $\sqrt{2}$.  For the incommensurate staggered square lattice we observe the preservation of the main two gaps of the butterfly for $\phi/\phi_0<0.5$, shown in Fig.~\ref{fig:HofsButtSquare}(a), much like the gaps that appear in Fig.~\ref{fig:HofsButt}(a) and \ref{fig:HofsButt8Fold}(a). When we consider a larger flux region, shown in Fig.~\ref{fig:HofsButtSquare}(b), we observe a similar quasiperiodic nature of the energy-flux plane to that of the quasicrystal. Therefore, we can confirm that the incommensurate nature of the different tilings of the standard quasicrystalline quasilattices plays an important role in the quasiperiodic nature exhibited in the Hofstadter butterflies.

\subsection{Finding topological states: The Bott index}

We next investigate the topology of the states. For periodic systems, non-trivial topology can appear in the form of a non-zero Chern number. In our case, however, we cannot calculate the Chern number directly, because we are not considering a periodic system, and the definition of a topological invariant is difficult. A related real-space quantity is the Bott index, which discerns between pairs of unitary matrices that can or cannot be approximated by a pair of unitary matrices that commute with them. It has been shown to be equivalent to the Chern number on a torus \cite{toniolo2017} and has been utilised in multiple scenarios \cite{Toniolo2018,zeng2019}, including for the quantum spin Hall effect in a Penrose quasilattice \cite{Huang2018a,Huang2018b}.

We will now define the Bott index. We scale the two-dimensional quasilattice by a constant factor, so that it fits inside a unit square. We denote the coordinates of the sites inside the unit square by $\tilde{x}_i\in[0,1]$ and $\tilde{y}_i\in[0,1]$. Given the two diagonal matrices $X_{i,j} = \tilde{x}_i \delta_{i,j}$ and $Y_{i,j} = \tilde{y}_i \delta_{i,j}$, we can define two unitary diagonal matrices
\begin{equation}
\hat{U}_x = \exp\left( i 2 \pi \hat{X}\right), \: \: \: \hat{U}_y = \exp\left( i 2 \pi \hat{Y}\right).
\end{equation}
We can also define a projector onto the eigenstates of up to energy $\epsilon$ as
\begin{equation}
\hat{P} = \sum_{\epsilon' < \epsilon} \ket{\epsilon'} \bra{\epsilon'},
\end{equation}
with $\ket{\epsilon'}$ the eigenstate of the system with energy $\epsilon'$. We can then project the unitary matrices into the eigenstates of upto energy $\epsilon$ and define two new matrices,
\begin{equation}
\hat{V}_x = \mathbb{1} - \hat{P} + \hat{P}\hat{U}_x \hat{P},
\end{equation}
\begin{equation}
\hat{V}_y = \mathbb{1} - \hat{P} + \hat{P}\hat{U}_y \hat{P}.
\end{equation}
Note, that $\hat{V}_x$ and $\hat{V}_y$ give the projected position operators. The Bott index can then be defined as \cite{Huang2018a}
\begin{equation}
\mathrm{Bott}(\epsilon) = \frac{1}{2\pi} \mathrm{Im} \mathrm{Tr} \log\left( \hat{V}_x \hat{V}_y \hat{V}_x^\dagger \hat{V}_y^\dagger \right),
\label{eq:Bott}
\end{equation}
where each projection is upto energy $\epsilon$. In addition, a numerically useful approximation of the Bott index is \cite{Toniolo2018}
\begin{equation}
\mathrm{Bott}(\epsilon) = \frac{1}{2\pi} \mathrm{Im} \mathrm{Tr}\left( \hat{P} \hat{U}_x \hat{P} \hat{U}_y \hat{P} \hat{U}_x^\dagger \hat{P} \hat{U}_y^\dagger \hat{P} \right) + O(1/N),
\label{eq:BottEst}
\end{equation}
with $N$ being the number of sites. Below, we will use both forms of the Bott index. The Bott index measures the commutativity of the projected position operators.

\begin{figure}[t]
	\centering
	\includegraphics[width=0.98\linewidth]{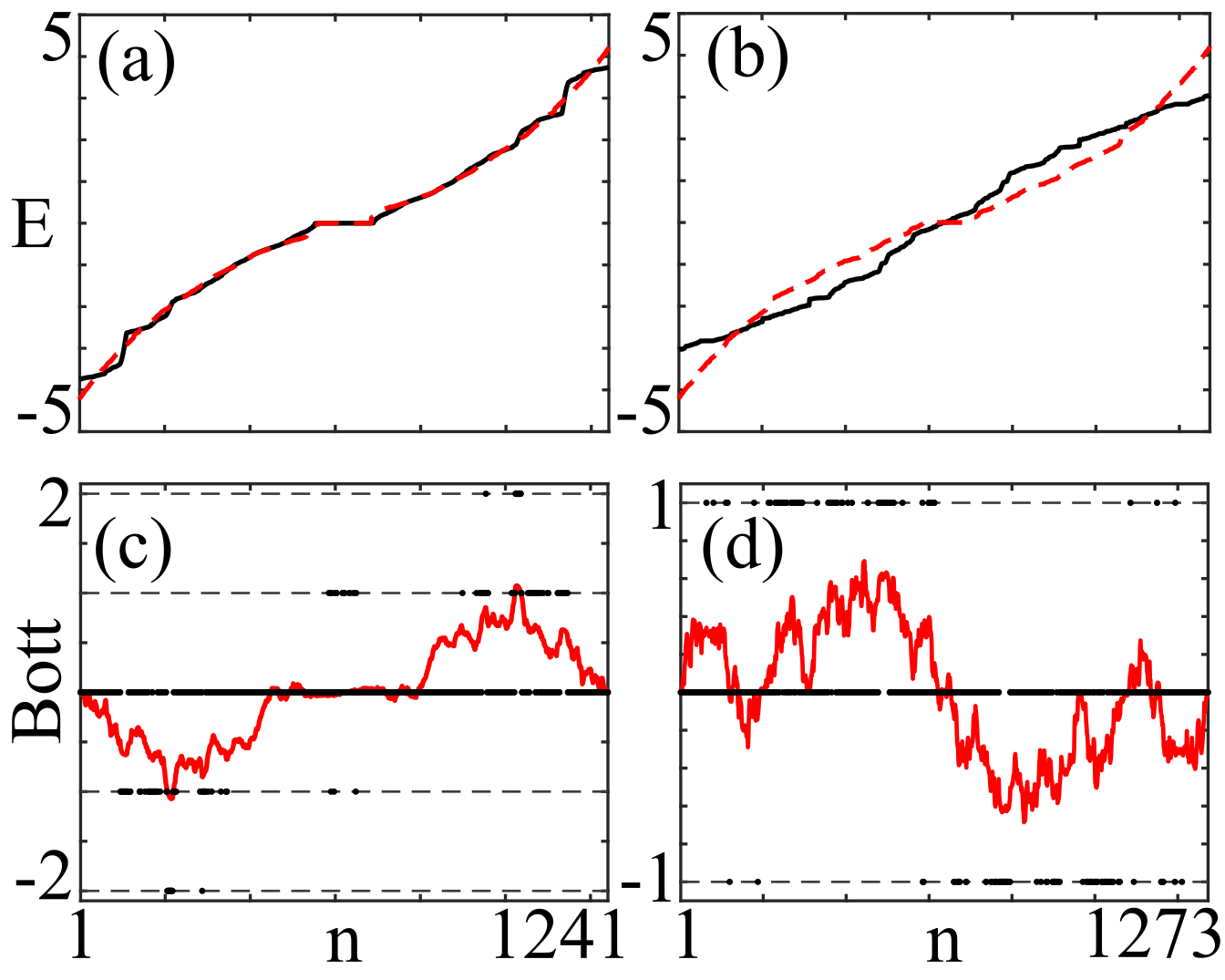}
	\caption{Vertex Hofstadter model for the Penrose five-fold ($N=1241$) and Ammann-Beenker eight-fold ($N=1273$) quasilattices. (a) The energy spectrum for the five-fold quasilattice with $\phi = 0$ shown by a solid (black) line and $\phi/\phi_0=0.1$ shown by a dashed (red) line. (b) Same as (a) but for the eight-fold quasilattice with $\phi = 0$ shown by a solid (black) line and $\phi/\phi_0=0.69$ shown by a dashed (red) line. (c) The Bott index shown with circles (black) and the Bott index estimate with a solid (red) line for the five-fold quasilattice and $\phi/\phi_0=0.1$. (d) Same as (c) but for the eight-fold quasilattice with $\phi/\phi_0 = 0.69$.}
	\label{fig:SpectraBott}
\end{figure}

To characterise the topological edge states of the vertex Hofstadter model we will look for states in the gaps of the energy-flux planes of Fig.~\ref{fig:HofsButt} and \ref{fig:HofsButt8Fold} with a corresponding non-zero Bott-index. We will take sizes of $1241$ sites for the five-fold Penrose quasilattice and $1273$ for the eight-fold Ammann-Beenker quasilattice. These sizes will allow for the states, or more strictly the probability densities, of the edge states themselves to be visualised in this paper. This becomes more difficult with larger system sizes. All results in this paper have been confirmed to be size independent by checking a variety of different system sizes. The Bott index is a relatively computationally intensive process and the decrease in system size is also beneficial, but not vital, to these computations.

We first consider two cuts of the Hofstadter butterflies of Sec.~\ref{sec:HofsButt} in Fig.~\ref{fig:SpectraBott}. The spectra for zero magnetic flux and a non-zero flux are shown in Figs.~\ref{fig:SpectraBott}(a) and (b) for each quasilattice. We have extensively checked various non-zero values of the magnetic flux and the shown examples exhibit features typical of each respective quasilattice. From the spectra it is immediately observed that the flat band for the eight-fold quasilattice is not as robust as that of the five-fold, due to the extra symmetries present in the eight-fold quasilattice. There is a small flat band of $<1\%$ of the states at $E=0$ for the eight-fold example, compared to $10\%$ of the states for the five-fold.  Note, this is not a result of the different flux used in each example, with the five-fold quasilattice showing a similar central flat band for $\phi/\phi_0=0.69$ and the eight-fold quasilattice not showing this for $\phi/\phi_0=0.1$.

We have calculated the Bott indices for the two non-zero magnetic flux examples, and they are shown for each state in Fig.~\ref{fig:SpectraBott}(c) and (d). As expected we do observe non-zero Bott indices, signalling that if the states where in a regular lattice then they would be in a band gap. There are multiple clusters of states with non-zero Bott indices shown in both spectra, due to many gaps opening in the system due to its non-periodic nature. There are also states in the five-fold quasilattice with larger than one Bott index, we have not been able to find these states in the eight-fold quasilattices at any investigated $\phi$ but we cannot rule out their presence somewhere due to the non-periodic nature of the flux. We also note that for the five-fold quasilattice the Bott index is non-zero for some of the states in the central flat band, see Fig.~\ref{fig:SpectraBott}(c). This is due to small gaps opening in the flat band of order $10^{-12}$, which is effectively zero compared to the order of the energy spectrum.

\subsection{Example edge states}

\begin{figure}[t]
	\centering
	\includegraphics[width=0.98\linewidth]{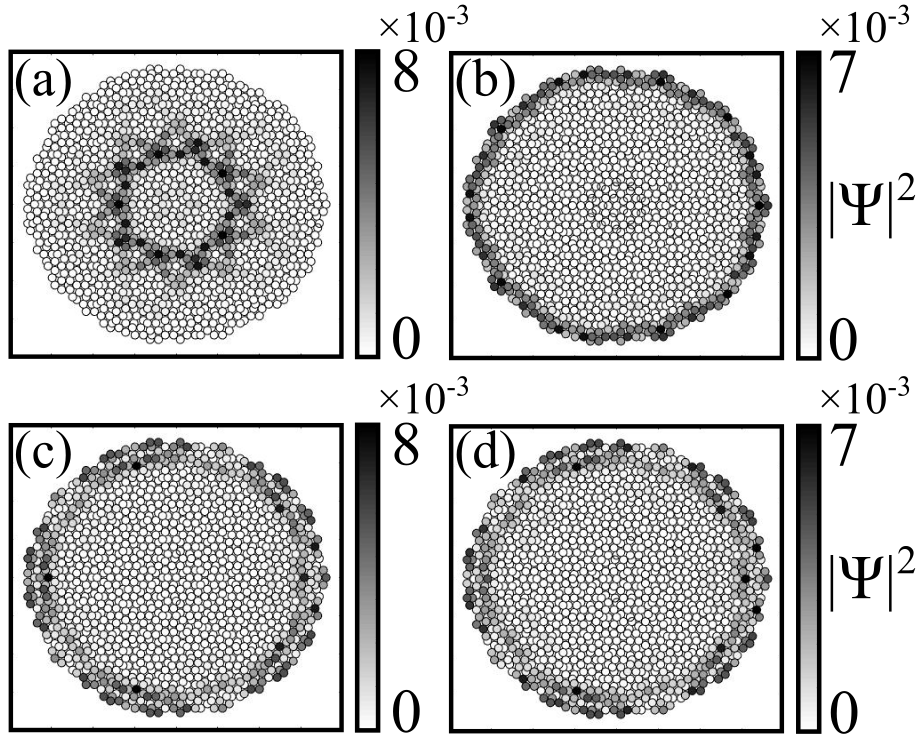}
	\caption{Example states (probability densities) for the Hofstadter vertex model on the five-fold Penrose quasilattice with $1241$ sites and $\phi/\phi_0 = 0.1$, corresponding to Fig.~\ref{fig:SpectraBott}(a) and (c). The example states are (a) the ground state, (b) state $107$ with Bott index $-1$, (c) state $213$ with Bott index $-2$, and (d) state $215$ with Bott index $-2$.}
	\label{fig:States5fold}
\end{figure}

We will now consider the structure of the states with non-zero Bott index in the two examples considered previously, i.e.\ for the five-fold we take $\phi/\phi_0=0.1$ in $1241$ sites and in the eight-fold we take $\phi/\phi_0=0.69$ in $1273$ sites. We have already confirmed that states with non-zero Bott index exist for these examples and that there is not one single gap with edge states in it for each spectrum but a set of gaps.

First, we will look at the case of the five-fold quasilattice, with four example states shown in Fig.~\ref{fig:States5fold}. The examples include the topologically trivial ground state, which is shown in Fig.~\ref{fig:States5fold}(a) and has an interesting structure which is five-fold symmetric. For $\phi/\phi_0=0.1$ each state with a Bott index of $\pm 1$ shows a similar form to Fig.~\ref{fig:States5fold}(b), which is the $107$th state with an energy of $E=-2.87$ and a Bott index of $-1$. This state takes a typically expected form of an edge state in a regular periodic lattice, with the state heavily localised to the edge. The edge state is five-fold symmetric, with a fine structure within each five-fold segment of the quasilattice due to the non-uniformity in each segment. We show in Figs.~\ref{fig:States5fold}(c) and (d) two example states of edge states with a Bott index of $-2$, and all states observed with a Bott index of $\pm 2$ in this example are of a similar form. These states have an interesting structure at the edges, with the state split between the edge and slightly away from the edge. The states with Bott index of $\pm2$, all occur within band gaps containing edge states of Bott index $\pm 1$, they are not contained within their own individual band gaps.

\begin{figure}[t]
	\centering
	\includegraphics[width=0.98\linewidth]{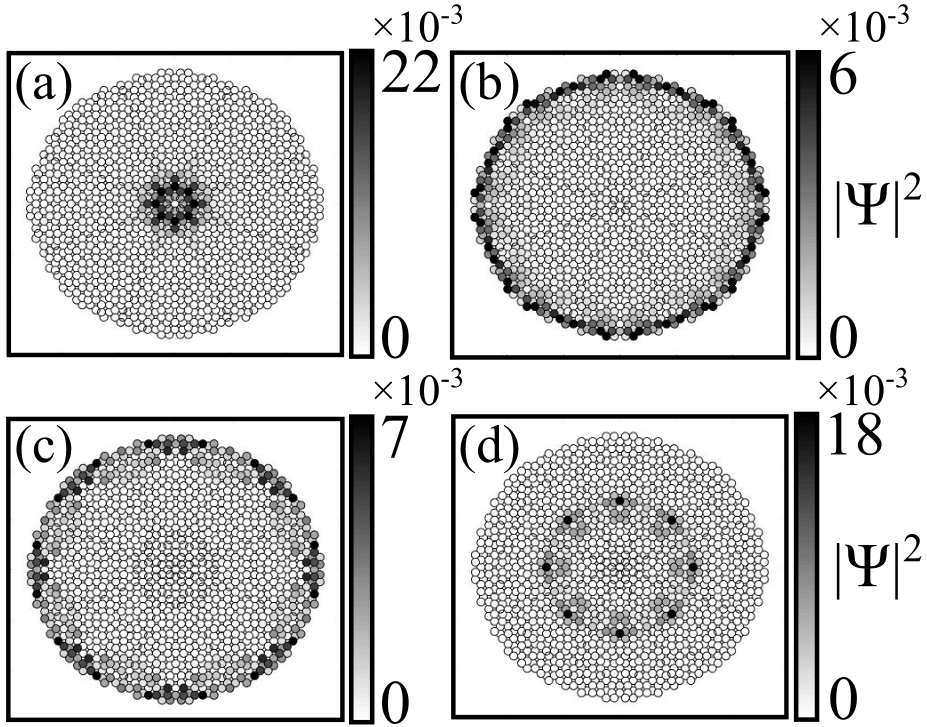}
	\caption{Example states (probability densities) for the Hofstadter vertex model on the eight-fold Ammann-Beenker quasilattice with $1273$ sites and $\phi/\phi_0 = 0.69$, corresponding to Fig.~\ref{fig:SpectraBott}(b) and (d). The example states are (a) the ground state, (b) state $246$ with Bott index $1$, (c) state $483$ with Bott index $1$, and (d) state $491$ with Bott index $1$.}
	\label{fig:States8fold}
\end{figure}

\begin{figure}[t]
	\centering
	\includegraphics[width=0.98\linewidth]{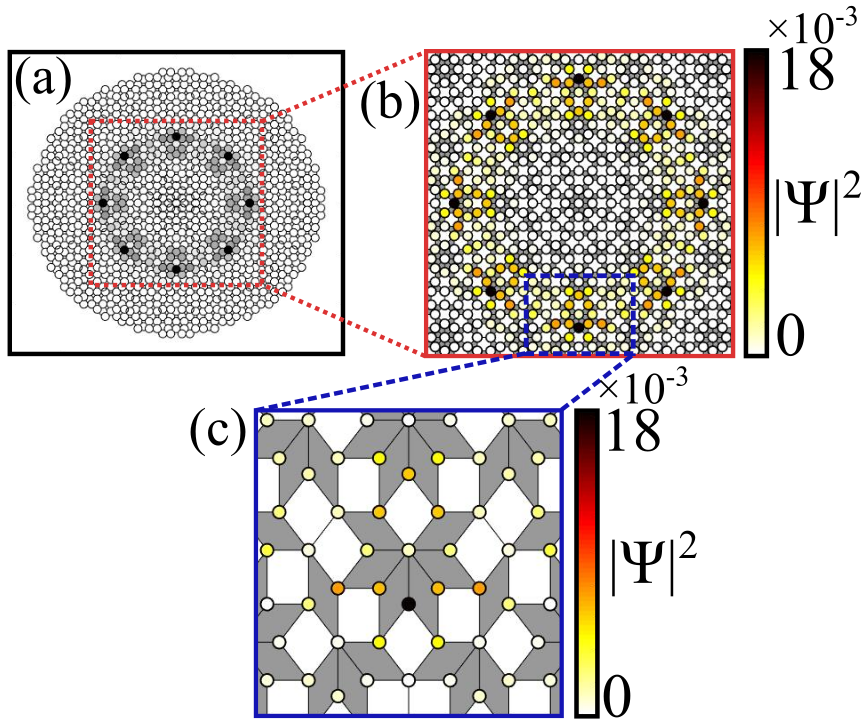}
	\caption{Focused plots of the internal edge state shown in Fig.~\ref{fig:States8fold}(d). (a) A repeated plot of the state. (b,c) Focused portions of the state showing its structure of being located on an almost full eight-fold star of the Ammann-Beenker tiling.}
	\label{fig:StatesInternal8fold}
\end{figure}

\begin{figure*}[t]
	\centering
	\includegraphics[width=0.8\linewidth]{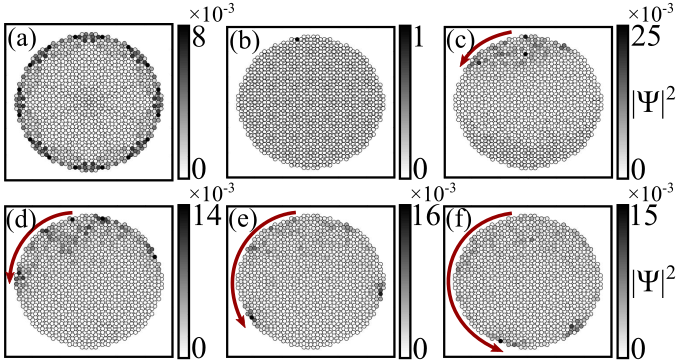}
	\caption{Exciting the edge states at the finite-size edge for the eight-fold Ammann-Beenker quasilattice with $1273$ sites for $\phi/\phi_0=0.69$. An example of the edge states on the finite size edge is shown in (a). The dynamics of a single site excitation on the edge is shown in (b-f) with (b) showing the state at $t=0\,J^{-1}$, (c) $t=25\,J^{-1}$, (d) $t=50\,J^{-1}$, (e) $t=75\,J^{-1}$, and (f) $t=100\,J^{-1}$.}
	\label{fig:ExternalDynamic}
\end{figure*}

\begin{figure*}[t]
	\centering
	\includegraphics[width=0.8\linewidth]{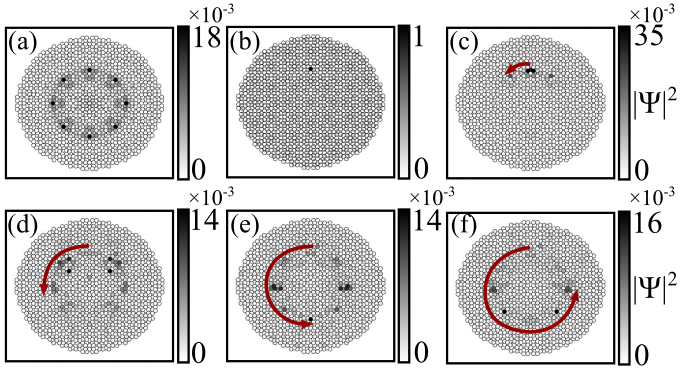}
	\caption{Exciting the edge states at the internal edge for the eight-fold Ammann-Beenker quasilattice with $1273$ sites for $\phi/\phi_0=0.69$. An example of the internal edge states in the bulk of the quasilattice is shown in (a). The dynamics of a single site excitation on the internal edge is shown in (b-f) with (b) showing the state at $t=0\,J^{-1}$, (c) $t=100\,J^{-1}$, (d) $t=200\,J^{-1}$, (e) $t=300\,J^{-1}$, and (f) $t=400\,J^{-1}$.}
	\label{fig:InternalDynamic}
\end{figure*}

We now turn to the eight-fold Ammann-Beenker quasilattice with four example states shown in Fig.~\ref{fig:States8fold}. Again, we have included the ground state in the examples, which is shown in Fig.~\ref{fig:States8fold} and contains the eight-fold symmetry. As shown in Fig.~\ref{fig:SpectraBott}(d) the Bott index for this example of the eight-fold quasilattice is either $0$ or $\pm 1$. We show two typical example edge states with Bott index $1$ in Figs.~\ref{fig:States8fold}(b) and (c). These states are eight-fold symmetric and localised to the edge of the system and have similar structural characteristics to the five-fold states of the form of Fig.~\ref{fig:States5fold}(b).

However, in the eight-fold quasilattice for various $\phi$ we observe a more exotic state, which has a Bott index of $1$ and is in a gap of the spectrum but is not localised to the edge of the finite-size system. Instead, this state is localised to an internal portion of the quasilattice, as is shown by the example of Fig.~\ref{fig:States8fold}(d). This states structure is intriguing and we consider it in more detail in Fig.~\ref{fig:StatesInternal8fold}. From the focused portions of the quasilattice, including the tiling, shown in Fig.~\ref{fig:StatesInternal8fold}, we can observe that the state is localised to an internal effective edge along almost complete eight-fold stars of the Ammann-Beenker tiling. Internal edge states have recently been found on fractal lattices \cite{Neupert2018,pai2019,Agarwala2019}. However, in the fractal lattice the internal edges are hard edges much like the outer edge (the finite size edge), whereas in the quasicrystal the internal edge states are not located on a hard edge. We therefore want to confirm that this internal edge state will indeed support transport, and we will investigate this next by considering the dynamics if we launch all our state into a site located on this internal edge. 

We have not found any internal edge states for the case of the five-fold Penrose quasilattice, but we cannot rule out their presence somewhere in the parameter space. For all fluxes where we have observed them, the set of internal edge states occur in the middle of a band gap and with regular edge states on either side. We have also investigated breaking the overall rotational symmetry of the lattice, and found the internal edge states to be independent of this. The exact nature of these states, why they occur and their topological properties are an open problem.

\subsection{Dynamics on the edge}

To consider the transport properties on the normal edge and the internal edge found in Fig.~\ref{fig:States8fold} we will consider the dynamics after the excitation of a single site on each edge. We will in this section consider only the eight-fold Ammann-Beenker quasilattice and will only note that similar behaviour for the five-fold edge states already discussed can be observed. We will consider the dynamics by initiating a state with all probability density in a single site and then evolve this under the Hamiltonian of Eq.~\eqref{eq:Hofs} for a small time step of $t = 0.005\,J^{-1}$. Since both positive and negative integer Bott indices are present in the spectrum there will be transport in both directions along the edge. Note, that we will also couple into other states which will result in an immediate spreading of the state into other sites, however, we would expect the main component of the state to exhibit transport along the edge.

First we will consider transport along the usual finite size edge as is shown in Fig.~\ref{fig:ExternalDynamic}. We excite a site at the edge and then evolve this with the resulting states all plotted in Fig.~\ref{fig:ExternalDynamic}. As expected, we observe the main component of the excitation moving along the edge in both directions. We have highlighted one of these directions in Fig.~\ref{fig:ExternalDynamic} with arrows to show the movement of the excitation around the edge.

In Fig.~\ref{fig:InternalDynamic}, we excite a single site on the internal edge found for the Ammann-Beenker quasilattice. We observe similar behaviour to the previous case of the finite size edge, with the main component of the excitation moving in both directions on this internal edge. We have again used arrows to highlight the transport in one of the directions. The timescale of the transport along this internal edge is considerably longer than that along the finite size edge. This is likely due to the structure of the internal edge states which appear to not be fully connected and are localised to regions along the internal edge as is shown by the example state shown in Fig.~\ref{fig:InternalDynamic}(a). The observation of the transport of an excitation along the internal edge is further evidence that it shows similar properties to the finite size edge even though it is not a hard boundary but a `soft boundary' in the quasilattice. The full nature of this internal edge state and the ability to predict its existence is a complex problem due to the quasiperiodic nature of the system and requires further study to understand.

\subsection{Flat bands with non-trivial topology}
\label{sec:Flat}

\begin{figure}[t]
	\centering
	\includegraphics[width=0.99\linewidth]{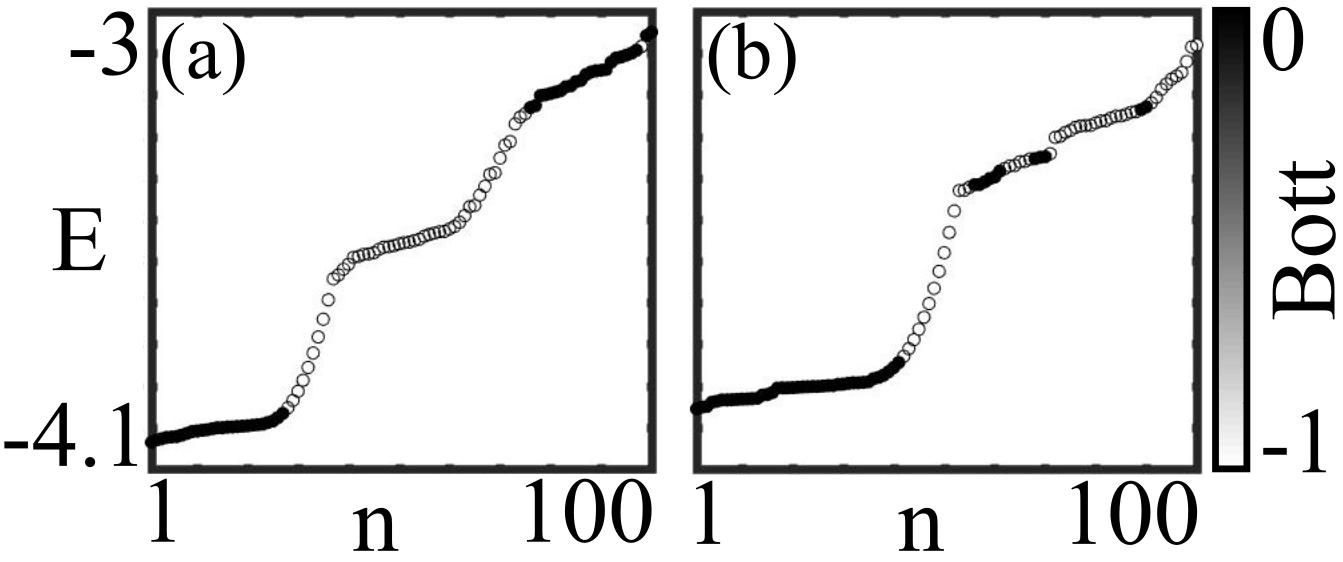}
	\caption{Examples of small topological flat bands in the Hofstadter vertex model for (a) the five-fold Penrose quasilattice with $1241$ sites and $\phi/\phi_0 = 0.034$ and (b) the eight-fold Ammann-Beenker quasilattice with $1273$ sites and $\phi/\phi_0=0.05$. The filling of the circles shows the value of the Bott index, with empty circles having a non-zero Bott index of -1 and filled circles having a Bott index of zero.}
	\label{fig:NoFlatBands}
\end{figure}

One method of generating topologically ordered states in lattices is to look for flatbands, or nearly flatbands, with non-zero Chern numbers which can show similar physics to Landau levels \cite{Neupert2011,regnault2011}. The idea being that when these nearly flatbands are integer or fractionally filled it is possible to realise the integer or fractional quantum Hall effect. We therefore now take a closer look at flatbands in the spectra.

The typical nearly flatband models considered are simpler in their spectrum than that of the quasicrystal. As we have observed already, the quasicrystalline quasilattice results in a spectrum with many gaps and it is difficult to search these for flatbands. Previous works have used a subset of quasicrystals with commensurate area tilings to observe Landau levels near band edges in the zero-field limit \cite{Fuchs2018}. In order to find some nearly flatbands with non-trivial topology we will also consider the limit of small field strengths and look at the band edges. These band edges are shown clearly in the Hofstadter butterfly of Figs.~\ref{fig:HofsButt} and \ref{fig:HofsButt8Fold} near the groundstate energies for low flux.

For small flux we can find lowest energy bands which are relatively flat for the system and contain about 2 to 4 percent of the total number of states in the spectrum. These states are, however, not nearly as flat as usually considered. The best case we find for the five-fold quasilattice is a ratio of bandwidth/band gap of $\approx 1/10$, which is shown in Fig.~\ref{fig:NoFlatBands}(a). For the eight-fold quasilattice the best example we find has a ratio of $\approx 1/4$ and is shown in Fig.~\ref{fig:NoFlatBands}(b). While the case found for the five-fold model might provide some hope for the flatband approach to topologically ordered states in quasicrystals, the behaviour of the spectrum for small variations in $\phi$ perhaps does not. When the flux is altered away from the considered value of Fig.~\ref{fig:NoFlatBands}(a) by approximately $1\%$ there can be the appearance of non-zero Bott index states which decrease the size of the gap and bring the ratio down to the order observed for the eight-fold model. The flatbands are hence not stable against small fluctuations in the magnetic flux.

Previous works on flatbands with non-trivial topology use multiple orders of tunnellings (next-nearest neighbours etc.). We find, however, that for the quasicrystals this makes the single particle behaviour more complex and using this we only find small flatbands of $<1\%$ of the states. Due to the complexity of the problem, it is not clear whether using an even higher level of fine tuning and/or adding more orbitals on each site could possibly lead to nearly flatbands with a non-trivial topology containing a large number of states, but the above computations already suggest that if such large flatbands exist, they are more difficult to obtain than for regular lattices.

\section{Topological order in models with interactions}\label{sec:Inter}

We now go beyond the non-interacting Hofstadter vertex model and construct topologically ordered models of fractional quantum Hall type on the five-fold Penrose and the eight-fold Ammann-Beenker quasicrystalline quasilattices. As detailed in the previous section, generating flatbands to obtain fractional Chern insulators does not look promising. Instead we use an approach, in which we start from an analytical fractional quantum Hall type wavefunction defined on the quasicrystalline quasilattice. Here, we shall consider the case of a Laughlin type state \cite{laughlin} with $1/q$ particles per flux, where $q$ is a positive integer.

Our starting point is a model derived in \cite{tu2014,nielsen2015,nielsen2018} for arbitrary lattices in two dimensions. The ground state of the model is the Laughlin type wavefunction
\begin{equation}
|\Psi_{Q}\rangle= \sum_{n_1,\ldots,n_N}\Psi_{Q}(n_1,\ldots,n_N)|n_1,\ldots,n_N\rangle
\end{equation}
with
\begin{multline}\label{L_State}
\Psi_{Q}(n_1,\ldots,n_N) \propto \delta_n \prod_{i<j}(z_{i}-z_{j})^{qn_{i}n_{j}}	
\prod_{i\neq j}(z_{i}-z_{j})^{- n_{i}} \\
\times \prod_{i,j}(w_i-z_j)^{p_in_j}.
\end{multline}
Here, $z_i=x_i+iy_i$ are the positions of the $N$ lattice sites making up the lattice, and $n_i\in\{0,1\}$ are the number of particles on each site. The particles are fermions for $q$ odd and hardcore bosons for $q$ even. The state also includes $Q$ anyons at the positions $w_i$ with charges $p_i/q$, where $p_i$ are integers. The $\delta_n$ factor fixes the number of particles to
\begin{equation}\label{particle_no}
M \equiv \sum_i n_i =  \frac{N - \sum_{k=1}^{Q}p_k}{q}.
\end{equation}
The difference between \eqref{L_State} and the original Laughlin state is that both the positions of the particles and the magnetic flux are restricted to be on the sites rather than being in a droplet-shaped region.

\begin{figure}[t]
	\centering
	\includegraphics[width=0.85\linewidth]{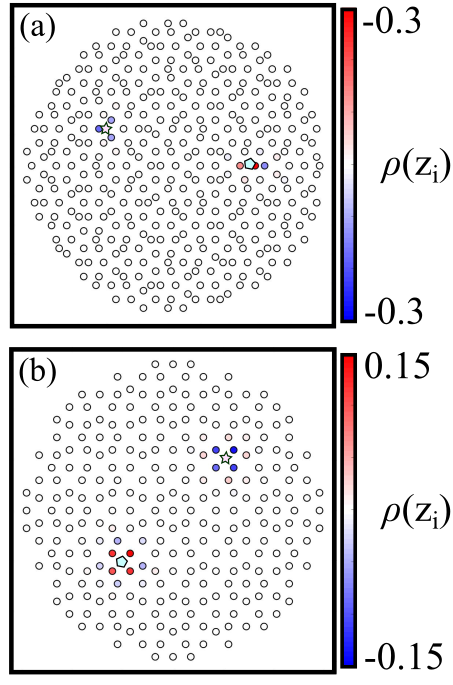}
	\caption{Density profiles $\rho(z_i)$ from Eq.~\eqref{density} on the (a) five-fold ($N=381$) and on the (b) eight-fold quasilattices ($N=273$) with one quasihole and one quasielectron in the system. The position of the quasihole is shown by a (red) star and the position of the quasielectron by a (blue) pentagon.}
	\label{fig:AnyonsLattice}
\end{figure}

It has been shown in \cite{nielsen2018} that the state \eqref{L_State} is the ground state of the Hamiltonian
\begin{equation}\label{Ham_4}
H=\sum_{i=1}^N\Lambda_i^{\dagger}\Lambda_i
\end{equation}
for $q + \sum_j p_j > 0$, where
\begin{equation}\label{Ham_L}
\Lambda_i = \sum_{j(\neq i)} \frac{\beta_j \hat{d}_j - \beta_i \hat{d}_i (q \hat{n}_j-1)}{z_i - z_j},
\end{equation}
and
\begin{equation}\label{Ham_L_op}
\beta_i = \prod_{j} (w_j - z_i)^{-p_j}.
\end{equation}
Here, $\hat{d}_j$ is the operator that annihilates a particle on site $j$ and $\hat{n}_j=\hat{d}_j^{\dag}\hat{d}_j$. This Hamiltonian consists of terms involving up to three sites.

\begin{figure}[t]
	\centering
	\includegraphics[width=0.99\linewidth]{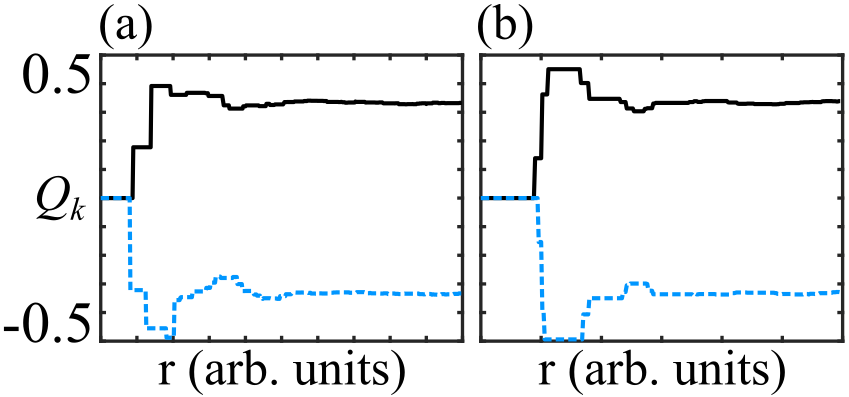}
	\caption{The excess charge distributions $Q_k(r)$ from \eqref{Excess_charge} plotted as a function of the radial distance $r$ from each anyon's position in (a) the five-fold and (b) the eight-fold quasilattices. We note that the anyons are well-screened and approach the right charges $\simeq \pm 1/3$. We find an error of $10^{-4}$ arising from the Monte Carlo simulations.}
	\label{fig:AnyonsCharge}
\end{figure}

We now consider this model on a quasicrystalline quasilattice by choosing the $z_i$ to be the vertices of the quasicrystal. Although the state \eqref{L_State} is reminiscent of a Laughlin state, it is not guaranteed that the state has the correct topological properties. We need to test this numerically. We do this by showing that the anyons are properly screened and have the correct charge and braiding properties.

Let us define the density profile of the anyons as
\begin{equation}\label{density}
\rho(z_i)= \langle \Psi_{Q \neq 0} | n_i | \Psi_{Q \neq 0} \rangle - \langle \Psi_{Q =0} | n_i | \Psi_{Q = 0} \rangle.
\end{equation}
This quantity measures how much the presence of the anyons change the expectation value of the number of particles on the site at $z_i$. If the anyons are properly screened, $\rho(z_i)$ is zero everywhere, except for sites within a small region around each of the anyons. We define the excess charge of the $k$th anyon to be
\begin{equation}\label{Excess_charge}
Q_k(r) = -\sum_{i=1}^N \rho(z_i) \Theta(r-|z_i-w_k|),
\end{equation}
where $\Theta(\ldots)$ is the Heaviside step function, i.e.\ the excess charge is minus the sum of $\rho(z_i)$ within a circular region of radius $r$. For properly screened and well-separated anyons, $Q_k(r)$ is constant, when $r$ is much larger than the size of the anyon, but small enough that the circular region is far from all other anyons in the system and far from the edge. This constant is called the charge of the anyon and should be equal to $p_k/q$.

We now test the screening and compute the anyon charges numerically for $q=3$. We insert one quasihole ($p_k=1$) and one quasielectron ($p_k=-1$) at the positions illustrated in Fig.\ \ref{fig:AnyonsLattice}. We compute $\rho(z_i)$ using Monte Carlo simulations, and the results (see Fig.~\ref{fig:AnyonsLattice}) show that the anyons are well-screened with radii of a few constants of the vertex model $l$. From Fig.~\ref{fig:AnyonsCharge}, we see that the charge is $\simeq \pm 1/3$ as expected.

It was shown analytically in \cite{nielsen2018} that the braiding properties of the state \eqref{L_State} are the same as for Laughlin anyons if the anyons are screened and well-separated throughout the braiding process. Given that we have already demonstrated screening in the systems, we hence also know that the braiding properties are as desired. 

The Hamiltonian \eqref{Ham_4} involves long-range interactions. It is likely, however, that one can truncate the Hamiltonian and still obtain a ground state with the same topological properties, as examples of this have been seen for regular lattices in two dimensions \cite{nandy2019}. Given that exact diagonalization for strongly correlated systems is limited to a few tens of sites, it is, however, very difficult to test this numerically.

\section{Conclusions}

In this paper, we have considered topological models in quasicrystalline quasilattices with rotational symmetry. We have focused on the five-fold Penrose and eight-fold Ammann-Beenker models, but expect our results to be indicative of the physics for quasicrystals with higher rotational symmetry. On each quasilattice we considered the Hofstadter vertex model, with the magnetic field introducing a phase factor upon tunnelling between sites. The incommensurate length scales of the magnetic field and the quasilattice resulted in the periodicity of the usual Hofstadter butterflies being destroyed. We found that a similar effect is observed in a standard periodic square lattice, when the flux is staggered along a single direction and the two fluxes are incommensurate with each other. This, along with the previous works on Rauzy tilings \cite{Tran2015,Fuchs2016}, showed that the destruction of the periodicity in the Hofstadter butterfly is due to the incommensurate length scales and not just the quasiperiodic nature of the quasilattice.

We have also studied the presence of topological states in both quasicrystals. We have found that standard edge states do occur, across a wide range of fluxes. There is also, in the case of the eight-fold quasilattice, some states with non-zero Bott indices that are not localised to the edge at the hard boundary. Instead these edge states are localised to a region within the bulk along the edge of some structures in the quasilattice. We confirmed that transport along both the standard edge and this soft internal edge occurs when a state localised to each is initialised. Interestingly, the internal edge states are not found in the five-fold quasilattice. The exact nature of the occurrence of the internal edge states and their prediction is an open problem. One approach to answer the generation of these states could be to consider the superspace picture in a similar manner to Ref.~\cite{valiente2019} but with the inclusion of a magnetic field.

We finished the paper by showing that topologically ordered models can also be obtained on the quasicrystalline quasilattices. Specifically, we investigated a model with an analytical Laughlin type ground state. We added anyons to the model and showed that they have the charge and braiding properties expected for Laughlin type anyons. We expect that a similar construction will also work for other fractional quantum Hall type states, such as Moore-Read states.

\begin{acknowledgments}
C.W.D. would like to acknowledge helpful discussions with Manuel Valiente and Dean Johnstone. C.W.D. acknowledges support from EPSRC CM-CDT Grant No. EP/L015110/1. This work was in part supported by the Independent Research Fund Denmark under grant number 8049-00074B.
\end{acknowledgments}

\appendix

\section{Derivation of phase factors on a general set of points}
\label{app:GeneralPhase}

As stated in the main text, a magnetic field being introduced to a charged particle on a lattice, or quasilattice, results in phase factors when hopping between sites. These phase factors take the form
\begin{equation}
\theta_{ij} = \int_{C_{ij}} \mathbf{A}(\mathbf{r}) \cdot d\mathbf{r},
\end{equation}
where $C_{ij}$ is the path from site $i$ to site $j$. We will consider the Landau gauge, which has a vector potential of $\mathbf{A} = B \left(0,x,0 \right)$. We will consider the phase introduced for a charged particle hopping between two sites of general position, $(x_i,y_i)$ and $(x_j,y_j)$. Using the known relations of line integrals of vector fields, we can write the phase factor from site $i$ to site $j$ as
\begin{equation}
\theta_{ij} = \int_0^1 \mathbf{A}(\mathbf{r}(t)) \cdot \mathbf{r}'(t) dt,
\end{equation}
with $\mathbf{r}'(t)$ the derivative of $\mathbf{r}$ with respect to $t$, where $0 \leq t \leq 1$ parametrises the path, and the path is given by
\begin{equation}
\mathbf{r}(t) = ( x_i(1-t) + x_j t, y_i (1-t) + y_j t, 0).
\end{equation}
Substituting in the form of the vector potential and using the defined path, the phase term is given by
\begin{equation}
\theta_{ij} = B (y_j-y_i) \int_0^1 dt \left[ x_i (1-t) + x_j t \right].
\end{equation}
Calculating the integral we obtain the general phase in the Landau gauge of a charged particle hopping between two sites $i$ and $j$ to be
\begin{equation}
\theta_{ij} = \frac{B}{2} (y_j-y_i) (x_i+x_j).
\end{equation}
If we consider a regular square lattice then we recover the expected phase term of
\begin{equation}
\theta_{ij} = Bl x_i,
\end{equation}
with $l$ being the distance between lattice sites and $x_i = x_j$.



%

\end{document}